\documentclass[prl,twocolumn,showpacs,superscriptaddress]{revtex4}%
\usepackage{amsfonts}
\usepackage{amsmath}
\usepackage{amssymb}
\usepackage{graphicx}%
\begin{document}

\title{First-principles study of native point defects in topological insulator Bi$_2$Se$_3$}

%\author{xxx}
%\affiliation{x}
%\affiliation{xx}
%\affiliation{xxx}
%\affiliation{xxxx}
%\affiliation{xxxxx}
%\affiliation{xxxxxx}
%\affiliation{xxxxxxx}

\author{Shuang-Xi Wang}
\affiliation{State Key Laboratory for Superlattices and
Microstructures, Institute of Semiconductors, Chinese Academy of
Sciences, P. O. Box 912, Beijing 100083, People's Republic of China}
\affiliation{Department of Physics, Tsinghua University, Beijing
100084, People's Republic of China}
\affiliation{LCP, Institute of
Applied Physics and Computational Mathematics, P.O. Box 8009,
Beijing 100088, People's Republic of China}
\author{Ping Zhang}
\thanks{Corresponding author; zhang\_ping@iapcm.ac.cn}
\affiliation{LCP, Institute of Applied Physics and Computational
Mathematics, P.O. Box 8009, Beijing 100088, People's Republic of
China}
\author{Shu-Shen Li}
\affiliation{State Key Laboratory for Superlattices and
Microstructures, Institute of Semiconductors, Chinese Academy of
Sciences, P. O. Box 912, Beijing 100083, People's Republic of China}


\date{\today}% It is always \today, today,

\begin{abstract}
The \emph{p}-type Bi$_{2}$Se$_{3}$ is much desirable as a promising
thermoelectric material and topological insulator, while the
naturally grown Bi$_{2}$Se$_{3}$ is always \emph{n}-type doped by
native point defects. Here we use first-principles calculations to
identify the origin of the \emph{n}-type tendency in bulk
Bi$_{2}$Se$_{3}$: The Se vacancies (V$_\text{Se1}$ and
V$_\text{Se2}$) and Se$_{\text{Bi}}$ antisite dominate the donorlike
doping with low formation energy, while the predisposed
Bi$_{\text{Se1}}$ defect results in the pair of V$_\text{Se1}$ and
Bi interstitial, which is also a donor rather than an acceptor.
Moreover, for Bi$_{2}$Se$_{3}$(111) surface, we find that the band
structures modulated by the defects explicitly account for the
existing experimental observations of \emph{n}-type preference.

\end{abstract}

\maketitle

Because of their excellent thermoelectric performance at room
temperature, the layered compound Bi$_{2}$X$_{3}$ (X=Se, Te), have
long been attracting great interest for decades in condensed-matter
physics \cite{Mahan1998,Kanatzidis2001}. Recently, these materials
have been drawing renewed attention by their novel properties as
topological insulators, which possess insulating gaps in the bulk
and gapless states on surfaces
\cite{Xia2009,Zhang2009,Chen2009,Tong2009}. Specially, the defects
in Bi$_{2}$Se$_{3}$ or Bi$_{2}$Te$_{3}$ dramatically dominate its
thermoelectric and topological properties. For Bi$_{2}$Te$_{3}$, it
can be simply made \emph{n} or \emph{p} type through variation in
the Bi:Te ratio. For Bi$_{2}$Se$_{3}$, however, one of the major
issues has been the difficulty in making the material \emph{p} type.
As revealed by angle resolved photoemission spectroscopy (ARPES),
the as-grown Bi$_{2}$Se$_{3}$ samples without doping are n-type
semiconductors \cite{Urazhdin2002,Urazhdin2004,Xia2009}, mainly
owing to the Se vacancies (V$_\text{Se}$ defects), or Se substituted
defects on Bi sites (Se$_\text{Bi}$ antisite defects). While the
acceptor-like defects such as Bi$_\text{Se}$ antisite defect, are
rarely observed. Only through doping with extrinsic Ca on the Bi
site, one can realize the formation of \emph{p}-type material
\cite{Hor2009}.

There have been several existing literatures dealing with the
defects in Bi$_{2}$Se$_{3}$
\cite{Hor2009,Urazhdin2002,Urazhdin2004}, while previous studies did
not investigate its defect structure and formation energy in detail.
Therefore a comprehensive studying of the native point defect
formation in this material is still lacking. Particularly, the
influence of the native point defects on the topological surface
states of this material is not much clear yet. In order to deeply
understand the novel properties of Bi$_{2}$Se$_{3}$, it is
instructive to investigate the formation of native defects, which is
not only of fundamental conceptual importance, but also paves the
way for realizing promising application of Bi$_{2}$Se$_{3}$ as a
novel topological insulator.

In this letter, by means of first-principles calculations, we
present a comprehensive study of native point defects in
Bi$_{2}$Se$_{3}$, including V$_\text{Se}$ vacancies, Se$_\text{Bi}$
and Bi$_\text{Se}$ antisite defects. The structure and formation
energy of such defects in bulk Bi$_{2}$Se$_{3}$, as well as their
band structures, are calculated and discussed in detail, confirming
the \emph{n}-type preference of Bi$_{2}$Se$_{3}$. Besides, by
analyzing the surface charge distribution and band structure of
Bi$_{2}$Se$_{3}$(111) surface with various defects, we present the
influence of native point defects on the topological surface states.

The calculations are performed within density functional theory
using the Vienna \textit{ab-initio} simulation package (VASP)
\cite{VASP}. The PBE \cite{PBE} generalized gradient approximation
and the projector-augmented wave potential \cite{PAW} are employed
to describe the exchange-correlation energy and the electron-ion
interaction, respectively. The spin-orbit interaction (SOI), which
has been confirmed to play an important role in the electronic
structure of the topological insulator, is included during the
calculation.

\begin{figure}
\begin{center}
\includegraphics[width=0.8\linewidth]{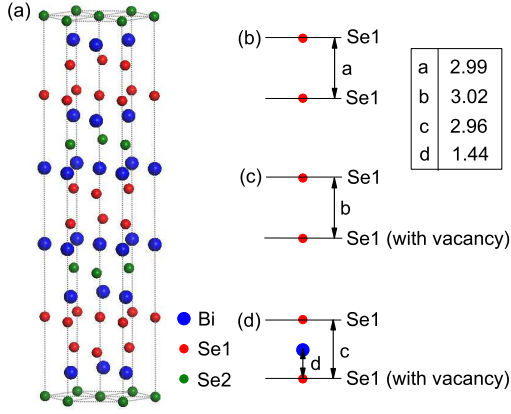}
\end{center}
\caption{(Color online) Atomic structure of bulk Bi$_2$Se$_3$ (a),
and inter-layer structure of perfect bulk (b), V$_{\text{Se1}}$ (c),
and Bi$_\text{Se1}$ (d).} \label{fig1}
\end{figure}

The crystal structure of Bi$_{2}$Se$_{3}$ is rhombohedral with the
space group $D_{3d}^{5}$($R\bar{3}m$). Equivalently, it can be
represented in terms of a hexagonally arranged layer structure,
which is shown in Fig. \ref{fig1}(a). The hexagonal unit cell has
three sets of quintuple layers, where each quintuple layer consists
of five atoms, forming a stable unit Se1-Bi-Se2-Bi-Se1 with strong
intrabilayer bonds, while the inter-layer bonding is much weaker.
Defect calculations are performed in a 3$\times$3$\times$1 supercell
with 135 atoms, and integration over the Brillouin zone is done
using the Monkhorst-Pack scheme \cite{Monkhorst1976} with
5$\times$5$\times$2 grid points. The cutoff energy for the plane
wave expansion is set to 400 eV, and the structure is fully
optimized until the maximum residual ionic force is below 0.02
eV/\AA. For the surface calculations, the slabs are modeled by a
6-quintuple layer with the defects adjacent to the surface. The
calculated structure parameters of perfect Bi$_{2}$Se$_{3}$ are
$a=4.198$ \AA, $c=29.643$ \AA, which are close to experimental data
$a=4.143$ \AA, $c=28.636$ \AA \cite{Nakajima1963}.

The formalism describing the formation energy of point defects in
materials has been well established, which depends on the chemical
potentials as well as the Fermi level for charged defects. The
ranges of chemical potentials for Bi and Se should be evaluated as
\begin{equation}
\mu_\text{Bi}^0+\frac{1}{2}\Delta{H_f}\leq\mu_\text{Bi}\leq\mu_\text{Bi}^0,
\label{Eqn1}
\end{equation}
\begin{equation}
\mu_\text{Se}^0+\frac{1}{2}\Delta{H_f}\leq\mu_\text{Se}\leq\mu_\text{Se}^0,
\label{Eqn1}
\end{equation}
where $\mu_\text{Bi}$ and $\mu_\text{Se}$ are chemical potentials
for Bi and Se in Bi$_{2}$Se$_{3}$, $\mu_\text{Bi}^0$ and
$\mu_\text{Se}^0$ are chemical potentials of pure Bi and Se,
respectively, and $\Delta{H_f}$ is the formation enthalpy of
Bi$_{2}$Se$_{3}$. The formation energy of a defect X in charge state
\emph{q} is defined as \cite{Zhang1991}
\begin{equation}
E_{\emph{f}}[X^q]=E[X^q]\pm\mu_{X}-E[\text{bulk}]+q(E_V+\Delta{V}+\varepsilon_F),
\label{Eqn2}%
\end{equation}
where $E[X^q]$ is the total energy of the system containing the
point defect $X$, $\mu_{X}$ is the elemental chemical potential with
a positive sign for vacancies and a negative sign for interstitial
defects. $E[\text{bulk}]$ and $E_V$ are the total energy and
valence-band maximum (VBM) of the perfect supercell, respectively.
The correction term $\Delta{V}$ is added to align the reference
potential in the defect supercell with that in the perfect bulk, and
$\varepsilon_F$ is the Fermi level referenced to $E_V$. Apparently
the defect formation energy is function of the atomic chemical
potential and Fermi level.

\begin{table}[ptbh]
\caption{Defect formation energies (in eV) in Bi$_2$Se$_3$ at the Se-rich limit
($\mu_{\text{Se}}$=$\mu_{\text{Se}}^{0}$) and Bi-rich limit ($\mu_{\text{Bi}}$=$\mu_{\text{Bi}}^{0}$),
respectively, at VBM ($\varepsilon_{\text{F}}$=0 eV).}%
\label{table1}
\begin{tabular}
[c]{ccccccc}\hline\hline \ Defect && Charge &&
$\mu_{\text{Se}}$=$\mu_{\text{Se}}^{0}$ &&
$\mu_{\text{Bi}}$=$\mu_{\text{Bi}}^{0}$\\\hline
\ && 0 && 0.57 && -0.59 \\
\ V$_{\text{Se1}}$ && +1 && 0.64 && -0.52 \\
\ && +2 && 0.68 && -0.48 \\
\hline
\ && 0 && 0.94 && -0.23 \\
\ V$_{\text{Se2}}$ && +1 && 0.99 && -0.18 \\
\ && +2 && 1.06 && -0.10 \\
\hline
\ && 0 && -0.14 && 2.77 \\
\ Se$_{\text{Bi}}$ && +1 && -0.13 && 2.78 \\
\ && +2 && -0.03 && 2.88 \\
\hline
\ && 0 && 1.54 && -1.37 \\
\ Bi$_{\text{Se1}}$ && -1 && 1.68 && -1.23 \\
\ && -2 && 1.85 && -1.06 \\
\hline
\ && 0 && 3.05 && 0.14 \\
\ Bi$_{\text{Se2}}$ && -1 && 3.06 && 0.15 \\
\ && -2 && 3.07 && 0.15 \\
\hline\hline
\end{tabular}
\end{table}

Table \ref{table1} summarizes the formation energy of native point
defects in various charge state in Bi$_2$Se$_3$ at the VBM
($\varepsilon_F$=0 eV) at the Se-rich limit
($\mu_{\text{Se}}$=$\mu_{\text{Se}}^{0}$) and Bi-rich limit
($\mu_{\text{Bi}}$=$\mu_{\text{Bi}}^{0}$), respectively. It is
noticeable that for a specific defect the differences in formation
energy between different charge states are very tiny, which means
that the defects in Bi$_2$Se$_3$ are not evidently charged.
Actually, by applying Bader analysis \cite{Bader1990}, we can
determine the ionic charges of Bi and Se atoms. The value of the
calculated Bader charge of the individual Bi, Se1 and Se2 atoms in
perfect Bi$_2$Se$_3$ are 0.95e, -0.55e and -0.80e, respectively. It
is seen that the differences in electronegativities of these atoms
are not large enough to induce charged defect states, quite
different from that in other typical semiconductor materials such as
ZnO \cite{zhang2001}. Therefore, in the following discussion we will
only focus on the neutrally charged defect states.

We can find from table \ref{table1} that at the Se-rich limit there
is only one stable defect, i.e., the Se$_\text{Bi}$ antsite that has
negative formation energy -0.14 eV, consist with the previous
experimental observation \cite{Urazhdin2002}. At the Bi-rich limit,
obviously the Se vacancies with negative formation energy are
stable, and V$_{\text{Se1}}$ (-0.59 eV) is more energetic favorable
than V$_{\text{Se2}}$ (-0.23 eV). Correspondingly, stronger
structural relaxation in the vicinity of V$_{\text{Se1}}$ can be
found compared with that of V$_{\text{Se2}}$.  For V$_{\text{Se1}}$
the inter-layer distance is enlarged to be 3.02 \AA (see Fig.
\ref{fig1}(c)) compared with 2.99 \AA (see Fig. \ref{fig1}(b)) of
perfect bulk. While surprisingly, the Bi$_\text{Se1}$ antisite
defect at the Bi-rich limit, which has been considered as the
\emph{p}-type defect thus difficult to form
\cite{Urazhdin2004,Lovett1977}, possesses of the lowest formation
energy -1.37 eV. As illustrated in Fig. \ref{fig1}(d), an outwards
relaxation of about 1.44 \AA ~for the substituted Bi atom with
respect to the Se1 layer is observed, therefore actually the
predisposed Bi$_\text{Se1}$ defect results in a pair of
V$_\text{Se1}$ and Bi interstitial.

\begin{figure}
\begin{center}
\includegraphics[width=0.8\linewidth]{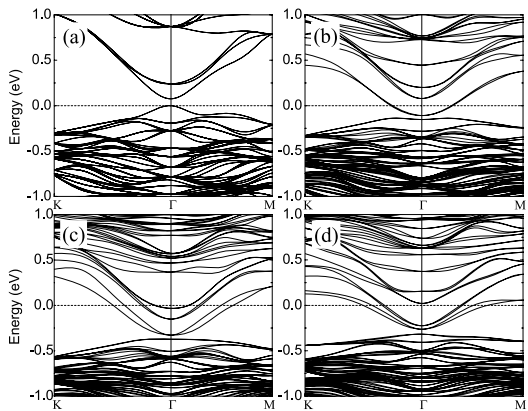}
\end{center}
\caption{Band structures of the bulk Bi$_2$Se$_3$: (a) for perfect
bulk, (b) with V$_{\text{Se1}}$, (c) with Se$_\text{Bi}$, and (d)
with Bi$_\text{Se1}$. The Fermi level is set to zero.} \label{fig2}
\end{figure}

In the following we present the electronic-structure results for
bulk Bi$_2$Se$_3$. Figure \ref{fig2} shows the electronic band
structure with and without point defects. For the perfect bulk, our
calculation yields a direct band gap of 0.08 eV at $\Gamma$ point
(see Fig. \ref{fig2}(a)), where the difference with the experimental
value of 0.2-0.3 eV \cite{Mooser1956,Black1957} may arise from the
deficiency of pseudopotentials adopted. When defects introduced, it
is clear that the gap remains robust, while the Fermi level of the
system undergoes an evident shift in energy, presenting somewhat
metallic behavior. For the bulk with defects V$_{\text{Se1}}$ and
Se$_\text{Bi}$, the Fermi level shifts up for 0.18 and 0.40 eV,
respectively (see Fig. \ref{fig2}(b) and (c)). These results are
consistent with previous conclusions that Bi$_2$Se$_3$ is naturally
\emph{n} doped \cite{Urazhdin2002,Urazhdin2004,Xia2009}. While for
the predisposed Bi$_{\text{Se1}}$ defect, the Fermi level shifts up
as high as 0.34 eV, implying that Bi$_2$Se$_3$ is also \emph{n}-type
doped. Keeping in mind the much low formation energy and the atomic
structure of this kind of defect mentioned above, it is much similar
to that in some II-VI semiconductuors \cite{Chadi1994,Chadi1999},
the pair of V$_\text{Se1}$ and Bi interstitial in Bi$_2$Se$_3$ forms
a stable donor complex (DX) rather than an acceptor.

\begin{table}[ptbh]
\caption{Defect formation energies (in eV) in Bi$_2$Se$_3$(111)
surface for neutral charge state at the Se-rich limit
($\mu_{\text{Se}}$=$\mu_{\text{Se}}^{0}$) and Bi-rich limit
($\mu_{\text{Bi}}$=$\mu_{\text{Bi}}^{0}$),
respectively, at VBM ($\varepsilon_{\text{F}}$=0 eV).}%
\label{table2}
\begin{tabular}
[c]{ccccccc}\hline\hline \ Defect &&&
$\mu_{\text{Se}}$=$\mu_{\text{Se}}^{0}$ &&&
$\mu_{\text{Bi}}$=$\mu_{\text{Bi}}^{0}$\\\hline
\ V$_{\text{Se1}}$ &&& 1.55 &&& 0.39 \\
\ Bi$_{\text{Se1}}$ &&& 2.57 &&& -0.34 \\
\ Se$_{\text{Bi}}$ &&& 0.35 &&& 3.26 \\
\hline\hline
\end{tabular}
\end{table}

Now that the basic properties of the native point defects in bulk
Bi$_2$Se$_3$ have been established. We may ask how these defects
affect its topological surface states, as the key important
properties of this topological insulator. There exists a rigid
difference of the chemical potential by about 200 meV between the
experimental result for the naturally \emph{n}-type sample and
theoretical calculation for the stoichiometric Bi$_2$Se$_3$(111)
surface \cite{Xia2009}. The physic behind this difference is
desirable to be investigated. Urazhdin \emph{et al.}
\cite{Urazhdin2002} have observed regular clover-shaped features on
Bi-doped Bi$_2$Se$_3$(111) surface, and attributed them to the
antisite Bi$_{\text{Se1}}$ at the bottom of the first quintuple
layer. Nevertheless, we believe that there must be other type of
defect on the surface. Based on the results about the defects in
bulk Bi$_2$Se$_3$, we calculate the atomic and electronic structures
of the surfaces with various defects, including V$_{\text{Se1}}$,
Se$_\text{Bi}$, Bi$_{\text{Se1}}$, as well as the perfect
Bi$_2$Se$_3$. In table \ref{table2} we list the formation energy of
the selected defects on Bi$_2$Se$_3$(111) surface for neutral charge
state. It can be seen that Se$_\text{Bi}$ at Se-rich limit (0.35 eV)
and V$_{\text{Se1}}$ at Bi-rich limit (0.39 eV) turn out to be
unstable compared with that in bulk Bi$_2$Se$_3$. While the defect
Bi$_{\text{Se1}}$ possesses the lowest formation energy (-0.34 eV)
at Bi-rich limit. Moreover, it is noticeable that for
Bi$_{\text{Se1}}$, unlike in bulk which forms a DX-like pair, the Bi
atom takes the position of Se1 atom, forming the surface antisite
defect Bi$_{\text{Se1}}$.

\begin{figure}
\begin{center}
\includegraphics[width=0.8\linewidth]{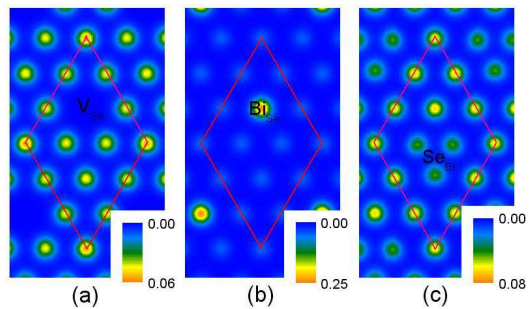}
\end{center}
\caption{(Color online) Surface charge density distributions (in
$\emph{e}/\text{\AA}^3$) at the height of 2.0 \AA ~above the
Bi$_{2}$Se$_{3}$(111) surface: (a) with V$_{\text{Se1}}$, (b) with
Bi$_\text{Se1}$, and (c) with Se$_\text{Bi}$.} \label{fig3}
\end{figure}

Figure \ref{fig3} presents the surface charge density distributions
of various defects near the surfaces. It is clear that there exists
an electron density depletion at the position of V$_{\text{Se1}}$
(see Fig. \ref{fig3}(a)). While for the Bi$_{\text{Se1}}$ (see Fig.
\ref{fig3}(b)), the substituted Bi atom on Se1 position can be
easily identified by the electron density accumulation. Because of
the low formation energies on (111) surface, and stabilities in bulk
Bi$_2$Se$_3$, these two kinds of defects should be easily observed
by STM experiment in Bi-doped sample. As for the Se-doped
Bi$_2$Se$_3$(111) surface, the Se$_\text{Bi}$ defect appears as a
regular clover-shaped feature (see Fig. \ref{fig3}(c)), with a clear
depletion of electron density compared with the perfect surface,
which can also be identified by experimental measurements.

\begin{figure}
\begin{center}
\includegraphics[width=0.8\linewidth]{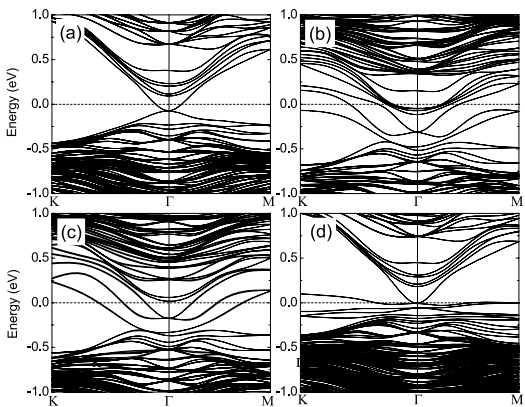}
\end{center}
\caption{(Color online) Surface band structure of the
Bi$_{2}$Se$_{3}$(111) surface: (a) clean surface, (b) surface with
V$_{\text{Se1}}$, (c) surface with Se$_\text{Bi}$, and (d) surface
with Bi$_\text{Se1}$. The Fermi level is set to zero.} \label{fig4}
\end{figure}

In Fig. \ref{fig4}, we plot the calculated band structures of the
Bi$_2$Se$_3$(111) surface with various defects. It can be seen that
for the perfect surface without any defect (see Fig. \ref{fig4}
(a)), the topological surface states form a single Dirac cone at the
$\Gamma$ point, implying non-trivial nature of this material.
Moreover, the Fermi level is -0.07 eV below the Dirac cone, hence
keeping far away from the bulk conduction band, which is much
different from the recent ARPES experimental observation that the
Fermi level resides in the conduction band \cite{Xia2009},
exhibiting naturally \emph{n}-type feature and presenting the
existence of intrinsic defects. To elucidate the experimentally
observed shifts of the Fermi level, let us investigate the band
structures of the systems with defects. For the V$_{\text{Se1}}$
(see Fig. \ref{fig4} (b)) and Se$_\text{Bi}$ (see Fig. \ref{fig4}
(c)), the Fermi level of the systems result in rigid shifts up about
0.24 and 0.10 eV compared with the perfect surface, respectively,
reaching into the bulk conduction band, hence presenting as donors.
Nevertheless, for Bi$_{\text{Se1}}$, the Fermi level almost overlaps
with the Dirac point, therefore it does not present evident
electron-donor or -acceptor feature. The difference with that in
bulk Bi$_2$Se$_3$ may arise from that here the substituted Bi atom
just takes the position of Se atom, rather than forming the
Bi-V$_{\text{Se1}}$ pair.

In conclusion, we have systematically studied the properties of
native point defects in topological insulator Bi$_2$Se$_3$. For bulk
Bi$_2$Se$_3$, V$_{\text{Se1}}$ and Bi$_{\text{Se1}}$ defects at the
Bi-rich limit were identified to be energetic stable. And the
existence of defects strongly modulate the band structures of
Bi$_2$Se$_3$. Moreover, we found that such defects have dramatically
influence on the surface states of this material, with
responsibility for the \emph{n}-type preference of Bi$_2$Se$_3$(111)
surface.

\end{document}